\documentclass[amsmath,amssymb,reprint,twocolumn,superscriptaddress,longbibliography, nobibnotes, prb]{revtex4-1}

\usepackage[english]{babel}
\usepackage[final]{changes}   
\usepackage[colorlinks=true, citecolor=blue, urlcolor=blue, linkcolor=blue]{hyperref} 
\usepackage{todonotes}

\usepackage{nicefrac}

\usepackage{multirow}
\usepackage{dcolumn}
\usepackage{bm}

\usepackage{graphicx}
\usepackage{xcolor}
\usepackage[export]{adjustbox}
\usepackage{float}
\usepackage{ulem} 

\usepackage{enumitem}
\usepackage{makecell}

\definecolor{DarkGreen}{rgb}{0.0, 0.5, 0.0}

\graphicspath{{Figures/}}

\newcommand{\TITLE}{Gradient-Based Optimization of Core-Shell Particles with Discrete Materials for Directional Scattering}

\begin{document}
	\title{\TITLE}
	\author{\firstname{Dalin} \surname{Soun}}
	\affiliation{LAAS-CNRS, Universit\'e de Toulouse, 31000 Toulouse, France}
	
	\author{\firstname{Antoine} \surname{Azéma}}
	\affiliation{LAAS-CNRS, Universit\'e de Toulouse, 31000 Toulouse, France}
	\affiliation{CEMES-CNRS, Universit\'e de Toulouse, 31000 Toulouse, France}
	
	\author{\firstname{Lucien} \surname{Roach}}
	\affiliation{Laboratoire de Chimie, CNRS, ENS de Lyon, 69364 Lyon, France}
	
	\author{\firstname{Glenna L.} \surname{Drisko}}
	\affiliation{Laboratoire de Chimie, CNRS, ENS de Lyon, 69364 Lyon, France}
	
	\author{\firstname{Peter R.} \surname{Wiecha}}
	\email[e-mail~: ]{pwiecha@laas.fr}
	\affiliation{LAAS-CNRS, Universit\'e de Toulouse, 31000 Toulouse, France}
	



	
	\begin{abstract}
		Designing nanophotonic structures traditionally grapples with the complexities of discrete parameters, such as real materials, often resorting to costly global optimization methods. This paper introduces an approach that leverages generative deep learning to map discrete parameter sets into a continuous latent space, enabling direct gradient-based optimization. For scenarios with non-differentiable physics evaluation functions, a neural network is employed as a differentiable surrogate model. The efficacy of this methodology is demonstrated by optimizing the directional scattering properties of core-shell nanoparticles composed of a selection of realistic materials. 
		We derive suggestions for core-shell geometries with strong forward scattering and minimized backscattering. Our findings reveal significant improvements in computational efficiency and performance when compared to global optimization techniques. Beyond nanophotonics design problems, this framework holds promise for broad applications across all types of inverse problems constrained by discrete variables.
	\end{abstract}
	
	\maketitle
	\section{Introduction}

	From the transmission of radio signals to the ongoing exploration of quantum computing -- the manipulation of electromagnetic waves has been a cornerstone of advancing technology ever since James Clerk Maxwell's groundbreaking work on electromagnetic fields.\cite{maxwellDynamicalTheoryElectromagnetic1865}
	In recent years, the field of nano-photonics has emerged as an exciting arena where the interplay of light and matter at the nanoscale can be controlled to an unprecedented extent.\cite{novotnyAntennasLight2011, kuznetsovOpticallyResonantDielectric2016}
	This control has powerful implications for applications ranging from information processing over sensing to subwavelength imaging. 
	However, the design of nanophotonic structures that can precisely dictate light behavior remains a challenging inverse problem, constrained by the limitations of available materials and intricate fabrication processes.\cite{elsawyNumericalOptimizationMethods2020, maDeepLearningDesign2020,bennetIllustratedTutorialGlobal2024}

	Traditional methods to address these inverse design challenges have relied on direct search and global optimization algorithms. These are computationally expensive and often converge very slowly due to the complex, high-dimensional search spaces involved.\cite{wuInvisibleDevicesNatural2022, teytaudDiscreteGlobalOptimization2022, majorelBioinspiredFlatOptics2024} Recent advancements in deep learning, particularly generative models, suggest a way forward by extracting complex correlations from large banks of data, and can also be trained to solve discrete inverse problems.\cite{liuTrainingDeepNeural2018, wiechaDeepLearningNanophotonics2021, jiangDeepNeuralNetworks2021, estrada-realInverseDesignFlexible2022, daiInverseDesignStructural2022, khaireh-waliehNewcomersGuideDeep2023} Such data driven methods however often suffer from insufficient dataset sizes or problematic data quality.\cite{wiechaDeepLearningNanophotonic2024} 
	
	Gradient-based optimization, either with adjoint formulations like in topology optimization, based on automatic differentiation, or in combination with deep learning surrogate models, has often been found to be the superior approach.\cite{jensenTopologyOptimizationNanophotonics2011, maBenchmarkingDeepLearningbased2022,baydinAutomaticDifferentiationMachine2018, hughesForwardModeDifferentiationMaxwells2019, minkovInverseDesignPhotonic2020, colburnInverseDesignFlexible2021}
	In combination with data-based physics predictors however, this requires very careful constraint handling, due to a high risk of erroneous extrapolation.\cite{dengNeuraladjointMethodInverse2021, khaireh-waliehNewcomersGuideDeep2023}
	Furthermore, these techniques generally do not work with discrete parametrizations, such as fixed sets of real materials, since gradients are then not defined. Complicated hybrid optimization procedures,\cite{hegdePhotonicsInverseDesign2020, kuhnInverseDesignCoreshell2022} customized interpolation schemes,\cite{zuoMultimaterialTopologyOptimization2017, liMultimaterialTopologyOptimization2018} or costly global optimization are typical fallback approaches.\cite{elsawyNumericalOptimizationMethods2020, bennetIllustratedTutorialGlobal2024}
	Recently, generative deep learning has been used  to re-parametrize image representations of structures to enable gradient-based optimization,\cite{augensteinNeuralOperatorBasedSurrogate2023, radfordInverseDesignUnitary2025}
	and it has been proposed to use the expressiveness of untrained neural networks by passing the original parametrization of a problem through a neural network and optimizing this network instead of the parameters.\cite{monakhovaUntrainedNetworksCompressive2021, chandrasekharMultiMaterialTopologyOptimization2021}

	In this paper, we propose an approach in a similar spirit, merging the strengths of automatic differentiation with deep generative models to address the design optimization problem of nanophotonic structures with discrete parameters. 
	We use a Wasserstein Generative Adversarial Network with Gradient Penalty (WGAN-GP) to map a discrete parametrization into a continuous latent space, enabling gradient-based techniques to probe the design space efficiently. 
	For situations where direct gradient evaluation is not feasible, a neural network can be used as a differentiable surrogate model,\cite{dengNeuraladjointMethodInverse2021} ensuring that the evaluation loop remains smoothly integrable\cite{khaireh-waliehNewcomersGuideDeep2023}.

	\begin{figure*}[t!]
		\centering
			\includegraphics[width=\linewidth]{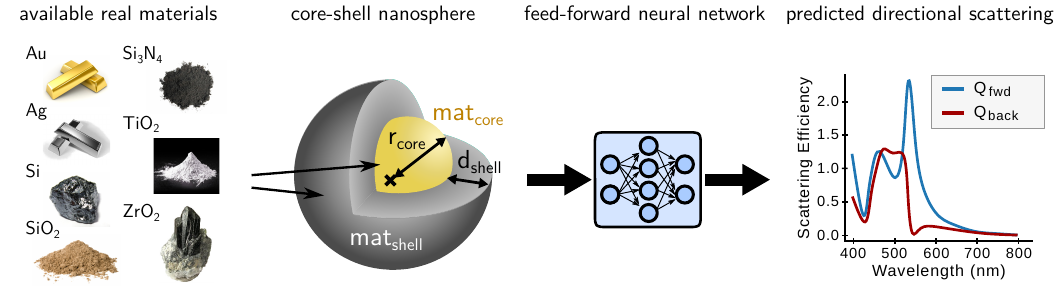}
		\caption{
			Illustration of the example problem of core shell nanopshere design with a discrete set of realistic materials. A feed-forward neural network model predicts the nanosphere's directional scattering response in form of the forward and backward scattering efficiencies $Q_{\textrm{fwd}}$ and $Q_{\textrm{back}}$ across the visible spectrum ($\lambda_0 =400$--$800$\, nm). 
			The inputs to the model are core material, shell material, core radius, and shell thickness.
		}
		\label{fig:forward_model_and_loss} 
	\end{figure*}

	Core-shell nano-spheres are a promising class of photonic structures, holding potential for applications in low-loss metasurfaces, sensing, quantum emitters or medicine  \cite{demarcoBroadbandForwardLight2021, lermusiauxSilverNanoshellsOptimized2023, shahCoreShellNanoparticleBased2014, sunCriticalRoleShell2020, kalambateCoreshellNanomaterialsBased2019, liuInverseDesignQuantum2023}. 
	Core-shell particles can be synthesized through bottom-up approaches, and are promising for scalable and cost-effective mass production.\cite{ghoshchaudhuriCoreShellNanoparticles2012, el-toniDesignSynthesisApplications2016}
	It is therefore compelling to design such structures for applications in nanophotonics.
	However, designing core-shell particles from a discrete selection of real materials, poses a challenge due to the lack of gradients for the discrete material parameters, which is why former approaches used simplifications like idealized permittivity,\cite{peurifoyNanophotonicParticleSimulation2018, dengNeuraladjointMethodInverse2021} less accurate direct deep learning methods,\cite{soSimultaneousInverseDesign2019} or complex optimization schemes to treat the discrete part separately.\cite{kuhnInverseDesignCoreshell2022}

	We demonstrate our approach of gradient-based discrete parameter optimization with the design of core-shell nano-spheres for optimal and strong forward light scattering.
	
	In this manuscript, we first introduce the details of our approach and its implementation for the optimization of core-shell nanoparticles. We describe the dataset generation, preprocessing techniques, and the architecture of the models involved. Subsequently, a statistical evaluation demonstrates the superior performance against traditional global optimization. Finally, we apply the method on the specific task of maximizing forward scattering whilst minimizing backscattering.
	\section{Gradient optimization with discrete parameters}

	\subsection{Problem: Design of core-shell nanospheres with discrete materials}

	Throughout this work, we illustrate our approach for gradient-based optimization of discrete problems by an example in nano-photonics design optimization.
	Our goal is to design spherical core-shell particles for light scattering. Specifically, we want to find particle geometries that maximize forward scattering while backscattering is minimized, which occurs at the so-called Kerker condition. This is highly interesting for applications in bottom-up fabricated, low-loss metasurfaces (so-called ``Huygens'' metasurfaces).\cite{kerkerElectromagneticScatteringMagnetic1983, deckerHighEfficiencyDielectricHuygens2015, rahimzadeganDipolarHuygensMetasurfaces2020, gigliFundamentalLimitationsHuygens2021}
	Our geometry is described by four parameters: core material, shell material, core radius, and shell thickness. 
	While the sizes of core and shell are continuous parameters, the respective materials are chosen from a list of available materials and represent discrete parameters.

	The optical scattering of spherical particles can be analytically solved by Mie theory.\cite{mieBeitrageZurOptik1908}
	The forward and backward scattering cross sections can thereby be expressed as function of the $n$th order Mie coefficients $a_n$ and $b_n$:\cite{liuGeneralizedKerkerEffects2018}

	\begin{equation}
		\sigma_{\textrm{fwd}} = \frac{\pi}{k^2} \left | \sum\limits_{n=1}^{\infty} (2n +1)(-1)^n (a_n - b_n) \right |^2
	\end{equation}
	
	\noindent
	and:
	
	\begin{equation}
		\sigma_{\textrm{back}} =  \frac{\pi}{k^2} \left | \sum\limits_{n=1}^{\infty} (2n +1) (a_n + b_n) \right |^2
	\end{equation}

	\noindent
	The Mie coefficients are obtained by matching the fields at the spherical interfaces between the different materials.\cite{bohrenAbsorptionScatteringLight1998}
	Finally, the scattering efficiencies $Q_{\textrm{fwd}}$ and $Q_{\textrm{back}}$ used here, are simply the cross sections divided by the geometric cross section $\sigma_{\textrm{geo}}=\pi r^2$ of the sphere with outer radius $r$.
	To calculate the directional scattering, we use a homemade, freely available Python / NumPy implementation ``pymiecs''.\cite{PeterWiechaPymiecs2024}
	
	The inverse problem however, is an ill-posed problem and generally cannot be solved directly.\cite{odomMultiscalePlasmonicNanoparticles2012, wiechaDeepLearningNanophotonics2021}
	We want to use gradient-based optimization to find solutions to the design problem.
	In addition to the discrete parametrization not being differentiable, also the Mie toolkit is not differentiable. Deriving symbolic derivatives of core-shell Mie theory is very tedious.
	Implementation in an automatic differentiation (AD) framework furthermore requires AD versions of spherical Bessel functions and their derivatives, which are so far not available in the most popular AD toolkits so far.
	Finally, in order to ensure proper gradient flow, non-discrete parameter passing to the Mie solver would need to be implemented, e.g. using interpolated mixing of the actual material properties.
	To avoid these challenges, we choose to replace the Mie solver by a feed-forward neural network, which is described in the following.

	\subsection{Data Generation and Preprocessing}

	We build a dataset using core radii between $1$\,nm and $100$\,nm and shell thickness between $1$\,nm and $100$\,nm. 
	The available materials are illustrated in figure~\ref{fig:forward_model_and_loss}, and comprise the plasmonic metals gold (Au) and silver (Ag), as well as the dielectric materials silicon (Si), silica (SiO$_2$), silicon nitride (Si$_3$N$_4$), zirconium dioxide (ZrO$_2$), and titanium dioxide (TiO$_2$).
	In addition to the radii and materials of core and shell, the scattering depends on the environment and on the illumination wavelength. 
	For the environment, we use vacuum ($n_{\textrm{env}}=1$). The illumination is at visible wavelengths from $\lambda_0 =400$\,nm to $\lambda_0 =800$\,nm, sampled with 64 equally spaced points.
	For every configuration, we calculate the scattering efficiencies of core-shell nanospheres in forward $Q_{\textrm{fwd}}$ and backward direction  $Q_{\textrm{back}}$ via Mie theory.
	In this way, we generate a total of 122500 data points.

	It is common in deep learning to preprocess and normalize the data before training the model, to ensure it is in a suitable format, and in a numerical range similar to a unit normal distribution.\cite{qamarDataPreprocessing2020}
	The categorical features (the core and shell materials) are represented in one-hot encoding (vectors of $N$ zeros, with the $i$-th element, corresponding to the represented category, being set to 1). The numerical features (the core and shell radii), are normalized using a min-max standard scaler \cite{goodfellowDeepLearning2016}, which re-scales the data into the range $[-1,1]$. 
	The numerical target variables (the forward and backward scattering efficiencies \(Q_{\textrm{fwd}}\) and \(Q_{\textrm{back}}\)) are preprocessed with a logarithm transformation (``log1p''), to reduce the strong skewness of the raw data distribution, followed by applying a min-max scaler. The inverse transformation is performed on the predictions of the feed-forward Mie neural network.
	Note, that we verified each step of the preprocessing procedure to improve training convergence. Specifically we found that ``log1p'' preprocessing does not only increase the global performance of the Mie predictor model cutting the residual average error roughly by half, it improves in particular weak scattering cases, in which cases relative errors improve significantly.

	\begin{figure}[t!]
		\centering
		\includegraphics[width=\linewidth]{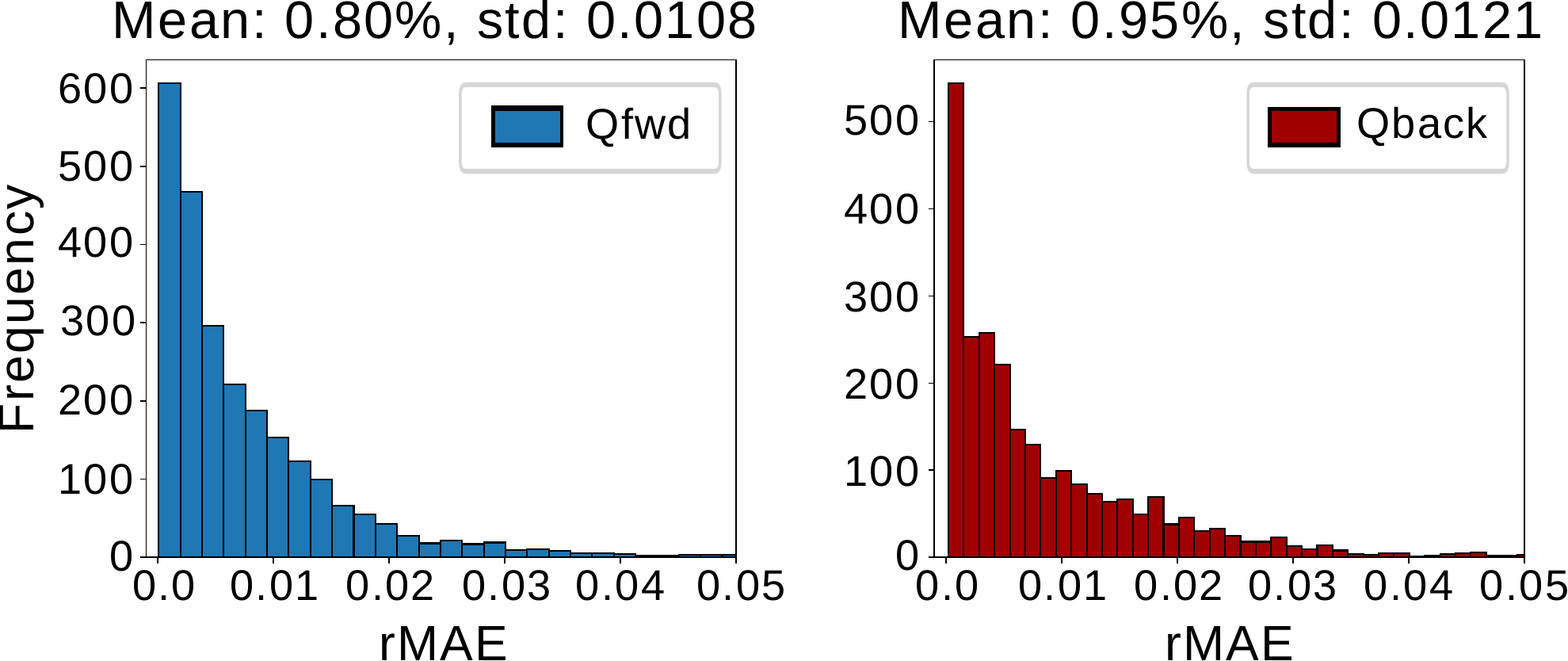}
		\caption{
			Relative Mean Absolute Error (rMAE) distribution for the \(Q_{\textrm{fwd}}\) and \(Q_{\textrm{back}}\) predictions from the feed-forward neural network on the test set. Note, that the histogram bars have different sizes because of different data ranges for forward and backward scattering.
		}		
		\label{fig:fwd_test_statistics}
	\end{figure}

	\subsection{Scattering modeling (the direct problem)}

	As mentioned above, gradient-based optimization requires all parts in the computational chain of scattering evaluation to be differentiable. 
	To this end, we replace the Mie toolkit in our demonstration by a learned feed-forward model.

	\paragraph*{Architecture of the scattering prediction model.}
	Our neural networks are implemented with Tensorflow.\cite{abadiTensorFlowLargeScaleMachine2015}
	The neural network takes the core-shell geometry as input and returns predictions for the scattering response (\(Q_{\textrm{fwd}}\)) and (\(Q_{\textrm{back}}\)). This is shown schematically in figure~\ref{fig:forward_model_and_loss}. 
	In our implementation we use a 1D residual network (ResNet), consisting of  two parts: 
	a first dense part takes as input the geometry and reshapes it into a 2D array for compatibility with 1D convolutional layers (Conv1D). 
	The second part consists of multiple 1D convolutional residual blocks and upsampling stages. Each residual block includes two Conv1D layers with batch normalization, leaky rectifier linear unit activations (leaky ReLU), and a skip connection.
	Sequences of four residual blocks are followed by a factor 2 upsampling layer, to gradually increase the data dimension towards the final output of $64$ wavelengths per spectrum.
	A final Conv1D layer consists of two output channels (one channel for \(Q_{\textrm{fwd}}\), the second for \(Q_{\textrm{back}}\)), followed by a linear activation function. 
	A detailed illustration of the scattering prediction model is shown in the supporting information figure~S2.
	
	Note that any differentiable model could be used instead of the ResNet. We found for example, that a simple multilayer perceptron (MLP), which evaluates a single wavelength instead of the full spectrum, also performs well, yet with errors approximately twice as large. The main advantage of an MLP would be, that it is not limited to the specific wavelengths defined by the dataset.

	\paragraph*{Training.}
	We use $108,000$ samples of our dataset for training and $12,000$ samples for validation and $2,500$ for testing. The model is trained using an MSE loss with the Adam optimizer\cite{kingmaAdamMethodStochastic2014} over 255 epochs, ended through early stopping after convergence (see supporting information figure~S3. 
	We use a hybrid approch \cite{smithDontDecayLearning2018} of increasing batch size up to 256 and a simultaneous learning rate decay schedule, which halves the learning rate every 15 epochs.
	The model converges well and training and validation loss curves show no sign of overfitting (see Sup.Info. Fig.~S3.
	
	\paragraph*{Performance.}
	The model performance is finally evaluated using a relative mean absolute error (rMAE) metric on the test set predictions. We find rMAE percentages as low as 0.80\% and 0.95\% for, respectively, \(Q_{\textrm{fwd}}\) and \(Q_{\textrm{back}}\). Histograms of the error distributions are shown in figure \ref{fig:fwd_test_statistics}.

	\begin{figure}[t!]
		\centering
		\includegraphics[width=\linewidth]{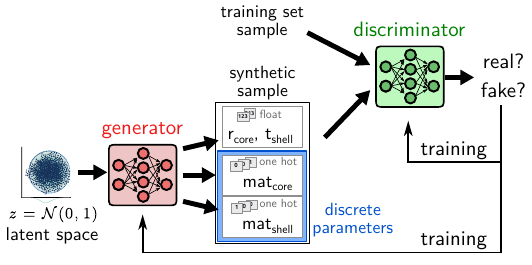}
		\caption{
		Global architecture of WGAN--GP Model. The generator network (red) takes inputs from a Gaussian distribution and produces two numerical values (core radius, shell thickness), and two categorical outputs (core material, shell material), the latter are returned as one-hot encoded vectors. 
		The discriminator network (green) is trained on distinguishing synthetic from real samples, and it is used as an optimized loss function for the generator training.
		}
		\label{fig:wgangp_model_and_losses} 
	\end{figure}

	\begin{figure*}[t!]
		\centering
		\includegraphics[width=\linewidth]{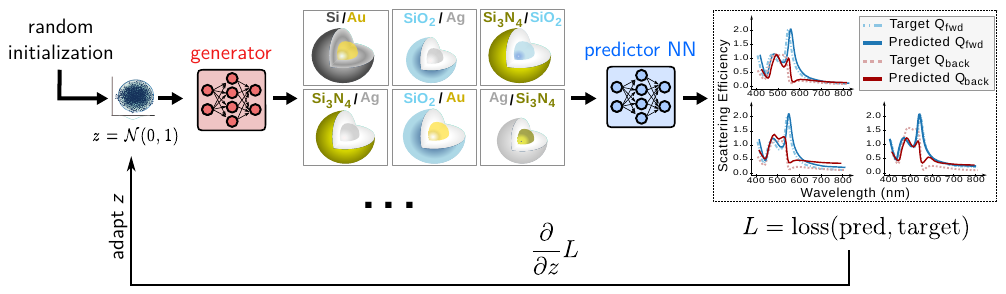}
		\caption{Our framework uses the pre-trained generator (red) and the feed-forward neural network (blue) for gradient calculation. The generator produces synthetic core-shell geometries from a set of randomly initialized latent vectors $z$. These geometries are then passed to the feed-forward model to predict their \(Q_{\textrm{fwd}}\) and \(Q_{\textrm{back}}\) spectra. 
		A loss function $L$ is defined to quantify the difference between predicted scattering and the target (this may for example be the mean square error between a full target reference spectrum and the predicted spectrum). The gradients of the loss $L$ with respect to the latent vectors $z$ are calculated via automatic differentiation and are used to adjust the latent vectors to better match the design target.}
		\label{fig:inverse_design} 
	\end{figure*}
	\subsection{Re-parametrization of Discrete Materials}

	Our goal is gradient-based optimization of the core-shell particles. Therefore, in addition to the differentiable surrogate for the scattering evaluation, we also need to convert the discrete material parameters of the geometry into a continuous representation.

	To do so, we employ a WGAN-GP,\cite{gulrajaniImprovedTrainingWasserstein2017, raghvendraNewRobustPartial2024} which is an adaptation of the original GAN\cite{goodfellowGenerativeAdversarialNetworks2014} with greatly improved training stability.
	GAN models such as the WGAN-GP learn a meaningful, smooth, and compact mapping of the original training data into a convex latent representation that follows a predefined distribution (we chose a standard normal distribution).\cite{sainburgGenerativeAdversarialInterpolative2019}
	The WGAN-GP has two main components. First the generator network (red in Fig.~\ref{fig:wgangp_model_and_losses}), which generates fake samples (here: core-shell geometries) from a latent vector input ($z$). 
	And second, the discriminator (green in Fig.~\ref{fig:wgangp_model_and_losses}). Its task is to distinguish whether data is fake (from the generator) or real (from the dataset), it is trained in alternance with the generator, for which it acts as a dynamically learned loss function.
	A detailed illustration of the WGAN-GP architecture we use, is shown in the supporting information figure~S2.

	Our WGAN-GP model is trained exclusively on the design parameters (sizes and materials of the core and shell), which are preprocessed in the same way as for the scattering prediction model.
	The architecture is based on two MLPs for generator and discriminator, in which each fully connected layer is followed by a leaky ReLU activation. 
	The generator uses an input latent dimension of 128, and has three distinct output layers: The first output layer consists of two neurons and ``tanh'' activation, yielding the numerical values for the core and shell radii.
	The second and third output layer use ``softmax'' activation, each with seven neurons (as here we consider seven distinct materials), yielding the materials for the core and the shell.
	This configuration guarantees that the WGAN-GP cannot generate unphysical geometries. It enforces the synthetic samples to lie within the data-normalization range, hence withing the user-defined parameters boundaries.
	Note that we deliberately use a large latent dimension, because information compression is not our goal. We require the smoothness and compactness properties of the latent space, together with the best possible generation quality.
	
	As training hyperparameters we found that a batch size of 256, and two different learning rates, $5\times 10^{-5}$ for the generator, and $5\times 10^{-4}$ for the discriminator optimizer (both Adam) give best results. 
	The other training parameters follow the original WGAN-GP paper.\cite{gulrajaniImprovedTrainingWasserstein2017}
	We train the WGAN-GP over 24 epochs.

	To assess the model fidelity, after the training, we generate $10,000$ samples and compare the distributions of the synthetic data with the uniform distribution of the training data, as shown in figure~S4.
	The distributions of the synthetic data's material parameters are almost uniform, meaning that the WGAN-GP model generates synthetic samples with the same statistics as the categorical features in the training set. The synthetic data distributions of the two size parameters are visually a little distorted, compared to the training set distributions. Yet, the statistics are still very similar and, most importantly, the entire training data parameter range is covered by the synthetic data. Therefore, an optimizer algorithm should be able to reach the full parameter range.

	\subsection{Gradient-Enabled Design Optimization}

	The actual design process consists in optimizing the core-shell nanostructures using gradient-descent methods to find a geometry that matches as closely as possible a target optical scattering response. 
	To this end, we define a fitness function that quantifies the deviation from the design target scattering response. This fitness function is either the mean square error with respect to a specific target scattering response, or simply the numerical value to be minimized or maximized. For maximization we use the negative of the value, as typical algorithms perform minimization by default.
	
	We connect the output of the pre-trained generator from the WGAN-GP to the input of the pre-trained Mie scattering prediction model. 
	The output of the scattering predictor is used for calculation of the design fitness. 
	The gradient of the design's fitness $L$ with respect to the generator's latent vector $z$ is calculated using automatic differentiation. $z$ is updated accordingly via a gradient-based optimizer (here we simply use again Adam). 
	The full workflow is depicted in figure~\ref{fig:inverse_design}.
	We use a learning rate of 0.01, which we found to be a robust and fast choice in a series of convergence test (see supporting information figure~S7). 
	Note that convergence speed could certainly be further optimized by employing a learning rate schedule and tuning the optimizer hyperparameters. 
	The latent representation $z$ is iteratively adjusted until convergence of the fitness.

	\begin{figure*}[t!]
		\centering
		\includegraphics[width=0.75\linewidth]{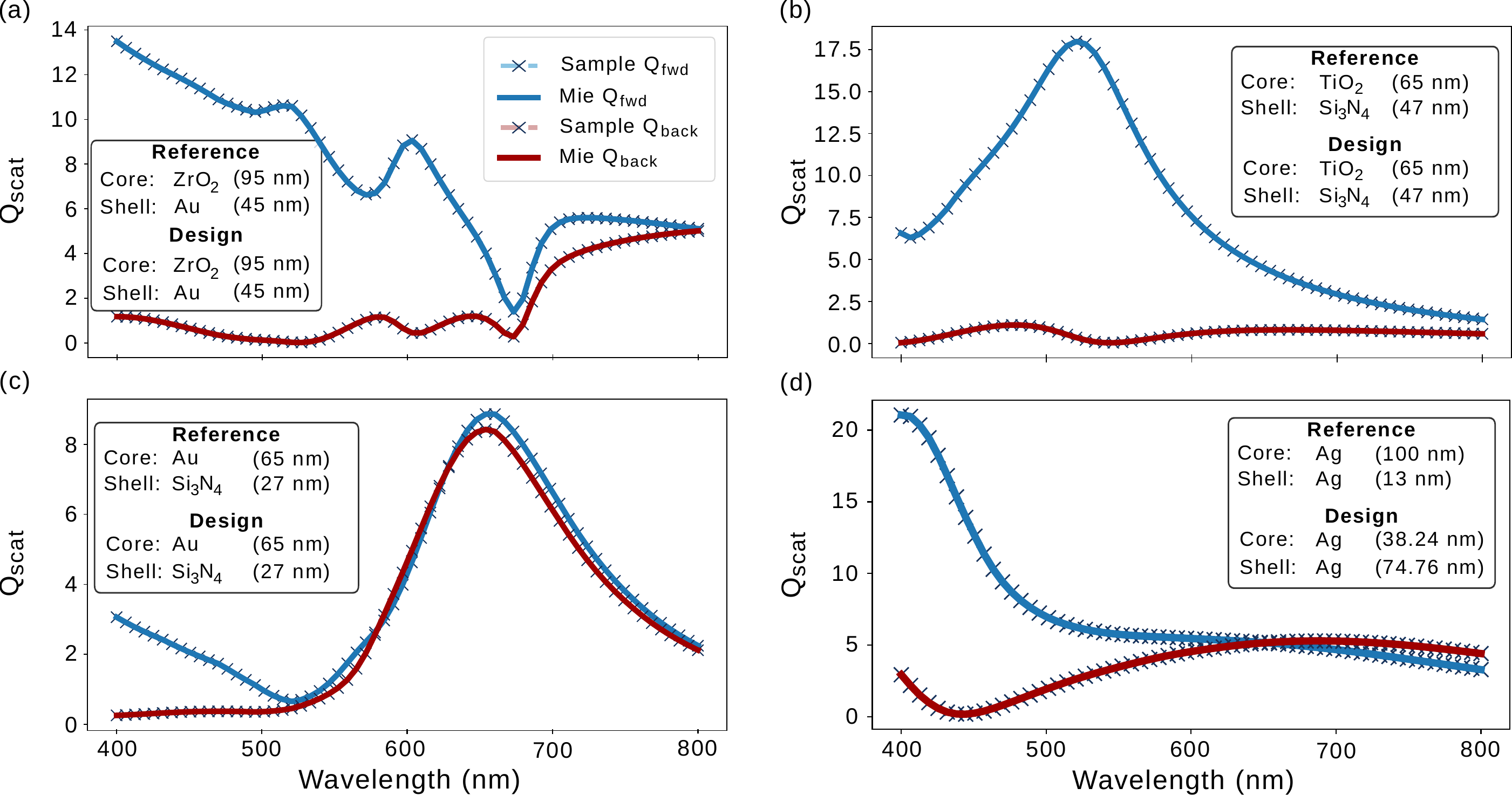}
		\caption{
			Random optimized scattering spectra vs the design targets. Each subplot (a-d) presents a different sample, showing both \(Q_{\textrm{fwd}}\) and \(Q_{\textrm{back}}\) for the design target and of the optimized geometries.
			The indicated numbers are core radius and shell thickness.
		}
		\label{fig:optimized_entire_spectrum}
	\end{figure*}

	A final problem needs to be resolved: gradient-based optimization converges towards the closest \emph{local} minimum. To nevertheless find solutions close to the global optimum, we initiate the optimization process with a large number of different, random latent vectors, which we then all optimize concurrently. 
	Thanks to the high degree of optimization of modern deep learning frameworks, this can be done with essentially no technical effort and batch optimization is extremely efficient. 
	On our system with an NVIDIA GeForce RTX 4090 GPU, optimization of 500 latent vectors in parallel for 100 iterations takes around 9\,s.
	For the here shown core-shell problem, 100 parallel optimizations were generally sufficient, we use 500 as it is not time-critical.  The required amount of initial guesses is typically problem specific. For problems with large numbers of local optima, gradient-free global pre-optimization with a gradient-based second step may be an efficient strategy.\cite{khaireh-waliehNewcomersGuideDeep2023}

	\section{Results}
	\label{sec:results} 

	\subsection{Examples and statistics from the test-set}

	\begin{figure}[t!]
		\centering
		\includegraphics[width=.95\linewidth]{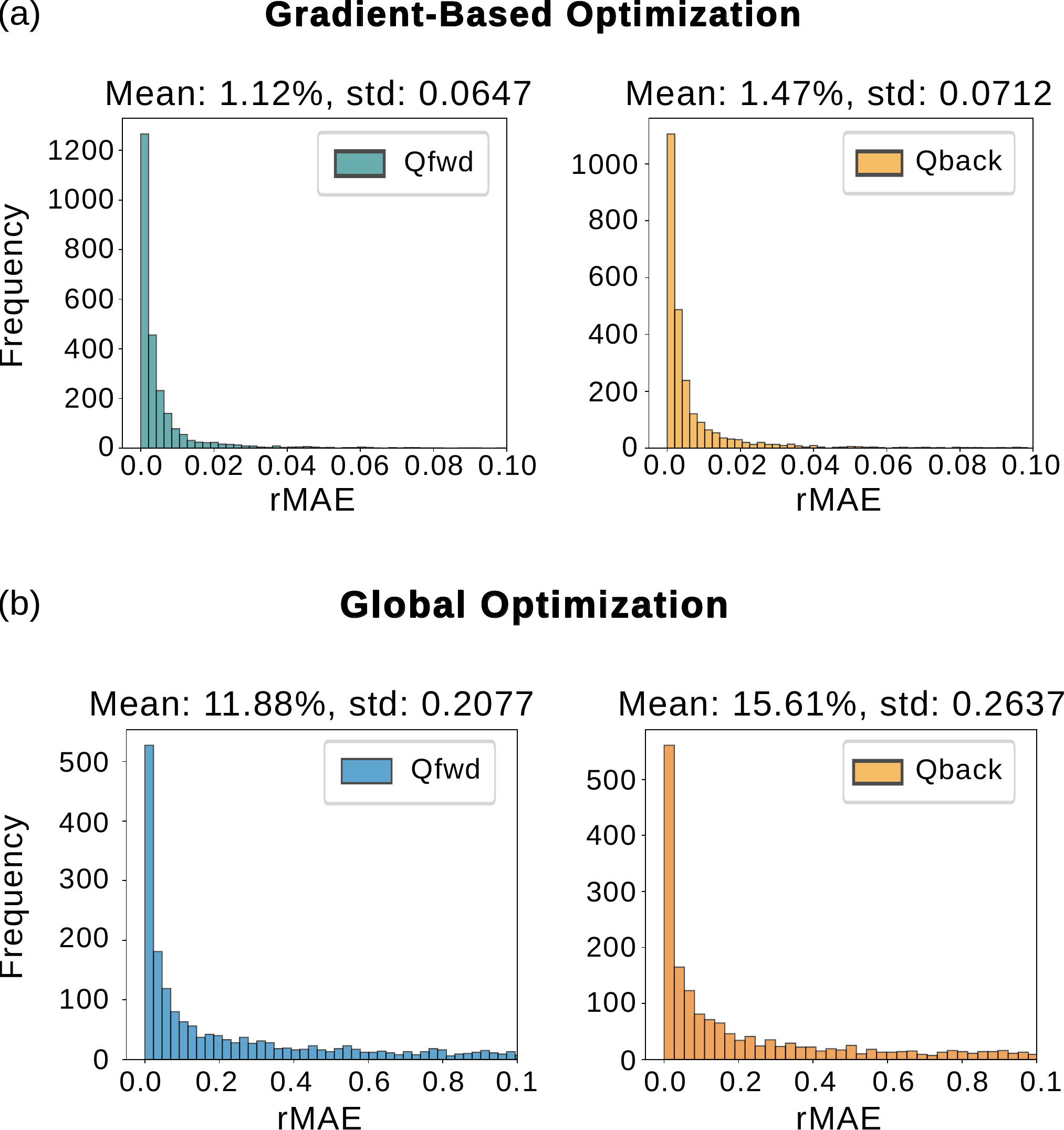}
		\caption{Comparison of relative mean absolute errors (rMAE) for \(Q_{\textrm{fwd}}\) (left) and \(Q_{\textrm{back}}\) (right).  
		(a) Gradient-based optimization.
		(b) Global optimization.
		}
		\label{fig:gradient_global_opt}
	\end{figure}
	
	We first test the performance of our optimization technique on the task of finding geometries from the test dataset based on the known optical spectra. The inverse design fitness function is the mean square error between the target spectrum and the predicted spectrum. Figure~\ref{fig:optimized_entire_spectrum} shows a selection of random examples, where the spectrum of the optimized geometry is recalculated with exact Mie theory. 
 	Remarkably, and despite the discrete material parameters, the gradient-based optimization reliably finds geometries that reproduce almost perfectly the desired solution.
	In figure~\ref{fig:optimized_entire_spectrum}d we also see that, as expected, ambiguous solutions (e.g. homogeneous particles with identical core- and shell-material) are not identically reproduced. 
 	The excellent performance of the optimization scheme is confirmed statistically on the entire test dataset (figure~\ref{fig:gradient_global_opt}a). We find a deviation of only around one percent between optimization results and the existing solutions.

 	To put this performance in context with conventional methods, we compare our gradient-based method to global optimization. For this comparison, we use the evolutionary optimization algorithm ``Differential Evolution'' (DE) from the ``NeverGrad'' toolbox to optimize the entire test dataset.\cite{simonEvolutionaryOptimizationAlgorithms2013, bennetNevergradBlackboxOptimization2021}
	DE is known to be a robust global algorithm that performs well on a large variety of problems.\cite{bennetIllustratedTutorialGlobal2024}

	100 iterations do not suffice to get reasonable results with global optimization. Therefore we let the global optimization run for 250 iterations, on a population of 100 individuals. 
	The average error with respect to the target is still around 10 times higher than with gradient-based optimization (mean absolute errors of around 1.12\% for gradient-based vs. around 11.88\% for global optimization). Please note also, that the latter uses an error-prone forward neural network for scattering predictions, while the global optimization uses exact Mie theory for fitness evaluation.
	We finally note, that global optimization with our vectorized Mie solver is similarly fast compared to the forward neural network framework. 
	However, the feed-forward neural network would be expected to have a drastic runtime advantage for computationally more difficult problems (such as free-form geometries), in additional to the strongly improved convergence rate.\cite{wiechaDeepLearningMeets2020}

	\subsection{Maximize forward scattering and minimize backscattering}	

	\begin{figure*}[t!]
		\centering
		\includegraphics[width=\linewidth]{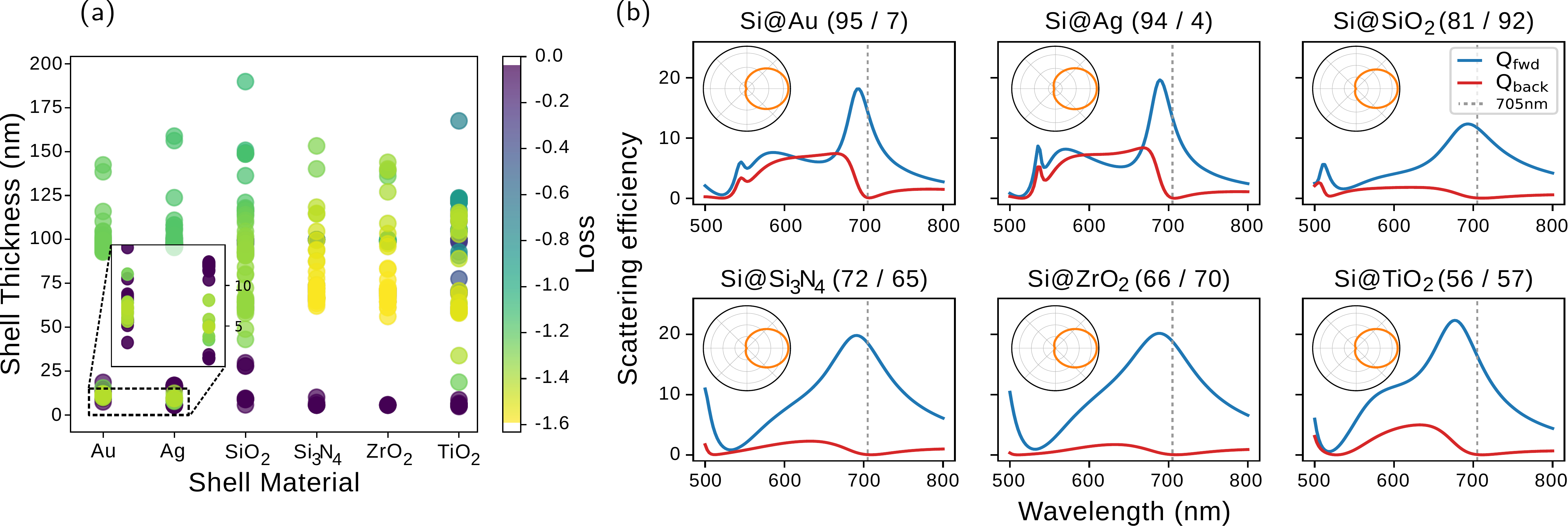}
		\caption{
			(a) Analysis of locally optimal regions in the design space for silicon core material nano-spheres. Scatter plots of 686 independent directional optimizations with weight $w=1$. The \(y\)-axis represents the shell thickness, the \(x\)-axis the shell material. The marker color corresponds to the optimization loss (lower is better).
			The inset shows a zoom on thin shell thicknesses for plasmonic shells.
			(b) Best design per shell material (fixed silicon core). The silicon core radius and shell thickness is indicated in parentheses above each subplot (in units of nm).
			Insets show the polar radiation patterns of the scattering at the design wavelength ($\lambda_0=705$\,nm).
		}
		\label{fig:design_space_plot}
	\end{figure*}

	
	Now we want to use gradient-based design optimization to find core-shell particles that achieve maximum forward scattering while minimizing backward scattering at specific wavelengths.
	This problem is more complicated than the above full spectrum matching, as it is effectively a multi-objective optimization.
	In earlier work we found that the quotient of forward over backscattering tends to mostly minimize backscattering \cite{wiechaDesignPlasmonicDirectional2019}. Therefore, to solve for strong forward scattering while minimizing backscattering, we define the following directional scattering fitness function: 
	
	\begin{equation}\label{eq:fitness_fwd_bwd_scat}
		L_{\text{dir}} = - Q_{\textrm{fwd}} + w Q_{\textrm{back}}\, ,
	\end{equation}
	
	\noindent
	which contains two terms that correspond to maximize forward and minimize backward scattering. To balance the two objectives, we add a weight $w$ to the backscattering term.
	We performed various series of random optimizations for different weights $w$. 
	These are described in the supporting information (see supporting information figure~S5). It turns out that a weight of $w=1$ is adequate for the goal of maximizing forward scattering with as little as possible backscattering, while not being too restrictive on the overall scattering efficiency.
	Note, that for each multi-objective optimization problem the fitness function should be chosen individually and, if applicable, its weight parameters should be optimized.
	We want to mention also, that there exist different ways to approach such multi-objective optimization (MOO) problems. The most popular alternative to a combined fitness function such as Eq.~\eqref{eq:fitness_fwd_bwd_scat} would be Pareto optimization, that searches all solutions that cannot be further improved without worsening at least one of the objectives \cite{wiechaEvolutionaryMultiobjectiveOptimization2017}. While Pareto-optimization historically uses gradient-free heuristics \cite{debMultiobjectiveOptimizationUsing2001}, currently research interests are emerging to develop gradient based MOO algorithms for machine learning applications \cite{chenGradientBasedMultiObjectiveDeep2025}. Our work may be of interest also in this context.

	Having set up the fitness function, we now use the possibility to perform very efficiently numerous optimizations with different initialization to globally explore the locally optimum regions in the design space \cite{dengNeuraladjointMethodInverse2021}, here solutions that maximize \(Q_{\textrm{fwd}}\), while minimizing \(Q_{\textrm{back}}\).
	Thanks to the WGAN-GP reparameterization, this is now possible also with discrete input parameters.
	We perform 5000 optimizations with random initializations, using a backscattering weight of $w=1$ and working wavelength at $\lambda_0=705$\,nm. During the optimization, we fix silicon as the core material, simply by replacing the WGAN-GP's output core material. After the optimization, we remove all non-silicon core samples, after this 686 solutions remain (which is approximately $1/7$, matching the 7 different available materials). Alternatively, a new WGAN-GP could have been trained for a fixed core material.

	We indicate in figure~\ref{fig:design_space_plot}a the fitness of the results by the color of the markers in a scatter plot. The color bar indicates the loss values after the optimization process, with yellow corresponding to the best (lowest) loss. The \(x\)-axis indicates the shell material, and the \(y\)-axis corresponds to the shell thickness. The core radius is also subject to optimization but is not shown in this figure. A detailed figure including core sizes is given in the supporting information figure~S6.
	
	Every dot corresponds to a local minimum in the optimization landscape. Therefore, ``gap'' regions with no solution indicate locally non-optimum parameter regions.
	This analysis reveals that two different types of locally optimal forward-scattering silicon core-shell nanospheres exist: (1) Structures with lower-index dielectric shell (SiO$_2$, Si$_3$N$_4$, ZrO$_2$, TiO$_2$) ideally have a large shell thickness while (2) plasmonic shells (Au or Ag) have an optimal thickness below $10$\,nm. In the case of silver, the shell should be even a bit thinner than for gold, presumably due to its better plasmonic properties and lower dissipation.
	The directional scattering spectra for the lowest loss solution of each shell material is shown in figure~\ref{fig:design_space_plot}b.
	These findings are in agreement with recent studies about such types of core-shell nanostructures \cite{oldenburgNanoengineeringOpticalResonances1998, tsuchimotoFanoResonantAlldielectric2016, demarcoBroadbandForwardLight2021}

	\section{Discussion}

	The optimized core-shell geometries predicted by our approach were chosen because of their feasible realization. Crystalline silicon cores in this size range have already been produced. \cite{chaabaniLargeScaleLowCostFabrication2019, eslamisaraySingleStepBottomupApproach2023, parkerUnveilingPotentialRedox2024} The creation of thin plasmonic shells made from gold (or even thinner silver) on a silicon core has not yet been produced. There are currently two reports of Si@Au core-shells in the literature, \cite{chaabaniSiAuCoreShell2021, sugimotoModeHybridizationSilicon2022} neither of which produced shells of suitable thickness or continuity to replicate the results predicted here. 
	
	However, thin plasmonic shells have been achieved to varying degrees of success on silica (SiO$_2$). \cite{shahCoreShellNanoparticleBased2014, oldenburgNanoengineeringOpticalResonances1998, lermusiauxSeededGrowthUltrathin2021}
	Silicon particles are typically coated by 2-3\,nm native silica layer, hence the surface chemistry is identical to a silica particle, allowing all techniques used on silica to be transposed to silicon. 
	A much thicker silica coating around a silicon core has already been achieved via thermal annealing in air \cite{tsuchimotoFanoResonantAlldielectric2016} and through the co-decomposition of precursors.\cite{demarcoBroadbandForwardLight2021} 
	Alternatively, it would be possible to deposit a silica shell around the silicon core using well-known sol-gel techniques.
	\cite{desertHighyieldPreparationPolystyrene2012, blasElaborationMonodisperseSpherical2008} Metal oxide cores have been coated with titania \cite{liCoreShellStructured2018, imhofPreparationCharacterizationTitaniaCoated2001} and zirconia \cite{arnalHighlyMonodisperseZirconiaCoated2006, kondratowiczNovelCuOcontainingCatalysts2019} shells using a variety of sol-gel protocols, rendering shells of comparable thickness to those studied here. 
	Thus, we selected core-shell structures not only that theoretically are optimized in terms of their forward/backward scattering ratio, but hold promise in terms of realizability.

	In this context of chemical feasibility, we can conclude from the above results that low-index dielectric materials seem the most interesting choice for fabrication of Huygens' sources with optimized scattering efficiency. 
	Silicon nitride and zirconium oxide seem to be particularly interesting, as in addition to strong directional scattering at the Kerker wavelength, the directionality is also very broadband. 
	The range of ``optimality'' over a wide span of shell thicknesses (figure~\ref{fig:design_space_plot}a) indicates also a certain robustness against fabrication size polydispersity.

	Alternatively, thin plasmonic shells made from gold or silver, seem to be capable of increasing the scattering efficiency at which the Kerker effect occurs. However, the experimental challenge of the fabrication of thin, yet crystalline and fully covering shells stills needs to be overcome.\cite{oldenburgNanoengineeringOpticalResonances1998}

	Please note, that all the solutions shown in figure~\ref{fig:design_space_plot} are locally optimal with respect to $L_{\text{dir}}$. This means they dependent on the choice of the backscattering weight $w$ in the optimization loss $L_{\text{dir}}$ and therefore do not necessarily represent ``physical'' feasibility.

	\section{Conclusions}

	In conclusion, we propose a WGAN-GP based re-parametrization of discrete problem parameters, to use gradient-based optimization on (partially) discrete problems. 
	To avoid local solutions, the (very fast) optimization can simply be run many times in parallel.
	Specifically we demonstrate the gradient-based design of directional optical scattering properties of core-shell nanospheres, specifically aiming to maximize forward scattering with as little backscattering as possible. 
	In comparison with global optimization, the gradient-based approach converges significantly faster. With a comparable compute budget, the error of the final solutions compared to the target, are in the order of 10\% for global optimization, while our gradient-based approach converges down to around 1\% residual error.
	Finally, we demonstrated by a practical example, how discrete parameter landscapes can be systematically analyzed for locally optimal solutions. 
	We found that silicon core spheres with dielectric shells seem to be very interesting candidates for broadband, strong directional scattering. 
	As a perspective, we foresee that gradient-based optimization will be very interesting for the design and study of nanophotonic multi-material structures and devices.
	The method is of course not limited to nanophotonics, it could be applied to any other discrete problem, possible examples may be synthesis protocol optimization in chemistry or process automation with binary triggers as system parameters.

	\section*{Disclosures}
	The authors declare no conflicts of interest

	\section*{Acknowledgments}
	We thank Aurélien Cuche for fruitful discussions.
	This work was supported by the French Agence Nationale de la Recherche (ANR) under grants ANR-22-CE24-0002 (project NAINOS) and ANR-23-CE09-0011 (project AIM), and by the Toulouse HPC CALMIP (grant p20010). 
	A.A. acknowledges funding by the Institute of Quantum Technology in Occitanie IQO (project Q-META).

	\section*{Data Availability Statement}
	All codes required to reproduce the results are available on a \href{https://github.com/S-Dalin/gradient_optimization_with_discrete_materials}{dedicated github repository}.
	
	\section*{Supporting Information}
	Supporting information can be found in a separate pdf document.

	\bibliography{2025_soun_discrete_materials_gradient_optimization.bbl}
	
\end{document}